\documentstyle[aps,12pt,manuscript]{revtex}

\begin{document}
\title{Neutron Halos in Heavy Nuclei \\ Relativistic Mean Field
  Approach\thanks{
  Supported by the Polish Committee of Scientific Research (KBN) under
  contract no. 2P03B04909} }
\author{Andrzej Baran, Krzysztof Pomorski \&\ Micha\l\ Warda}
\address{Institute of Physics, Maria Curie--Sk{\l}odowska University, \\
 20-031 Lublin, Poland}
\date\today
\draft
\maketitle

\begin{abstract}
Assuming a~simple spherical relativistic mean field model of the
nucleus, we estimate the width of the antiproton--neutron annihilation
($\Gamma_n$) and the width of antiproton--proton ($\Gamma_p$)
annihilation, in an antiprotonic atom system.  This allows us to
determine the halo factor $f$, which is then discussed in the context
of experimental data obtained in measurements recently done on LEAR
utility at CERN.  Another quantity which characterizes the deviation of
the average nuclear densities ratio from the corresponding ratio of the
homogeneous densities is introduced too.  It was shown that it is also
a good indicator of the neutron halo.  The results are compared to
experimental data as well as to the data of the simple liquid droplet
model of the nuclear densities. The single particle structure of the
nuclear density tail is discusssed also.
\end{abstract}

\section{Introduction}

The LEAR measurements at CERN~\cite{jas+93,lub+94}, show large neutron
halo factors for selected nuclei.  The observations of this factor were
performed by counting those nuclei which were born in annihilation
reactions in antiprotonic atoms. The following processes were observed:
$$\bar p + _Z^AX_N \longrightarrow _{Z-1}^{A-1} X _N + \{\pi\}\,,$$
$$\bar p + _Z^AX_N \longrightarrow _{~~Z}^{A-1} X _{N-1}+\{\pi\}\,.$$
Here, $_Z^AX_N$ denotes the nucleus with the atomic number $Z$, the mass
number $A$ and the number of neutrons $N$. The annihilation products
consist of a~few (usually 4-5) pions \{$\pi$\}.

In reactions shown the resulting nuclei have the number of neutrons or
the number of protons diminished by one as compared to the mother
nucleus.  Counting the annihilation products resulting from nuclear
neutrons ${\cal N}_n$ and from nuclear protons ${\cal N}_p$, one can
calculate the following halo factor $f=(Z/N){\cal N}_n /{\cal N}_p$.
For the case of "cold" annihilation reactions which take place at the
peripheral region of the nucleus and do not excite the final nuclei,
the factor $f$ is supposed to depend strongly on the ratio of neutron
and proton densities.

In order to describe the phenomenon of annihilation of the antiprotonic
atomic state on the mother nucleus we have done the calculation of the
halo factor $f$ and the density ratios deviation $h$, which we shall
define later. As a~basis of our calculation serves the relativistic
mean field model in which the field equations are solved numerically.
These solutions are exact even far outside the nuclear surface. This
precision is critical for the calculation of the quantities of
interest.  Both quantities $f$ and $h$, have been considered for
nuclei: $^{58}$Ni, $^{96}$Zr, $^{96}$Ru, $^{130}$Te, $^{144}$Sm,
$^{154}$Sm, $^{176}$Yb, $^{232}$Th, $^{238}$U.  All these nuclei were
studied in Ref.~\cite{jas+93,lub+94}.  The halo factor $f$ shows also
some dependence on the separation energies of the last nucleons (proton
or neutron).

This paper is organized as follows. In the next section we define and
discuss the model independent functions $f$ and $h$ which are supposed
to characterize halo phenomena.  An example, the droplet model
calculation of $h$ is shown.  The third section contains a~short
presentation of the relativistic mean field model which we have used to
determine $f$ and $h$ factors.

In the Summary of the paper we show and comment on the results of our
calculations

\section{Basic definitions}

In this section we define basic quantities which are suitable in
discussing halo phenomena. The first subsection collects the
approximations made in halo factor definitions. The second subsection
introduces new quantities. These are average densities calculated from
the density distribution and the deviation of average densities ratio
from the homogeneous neutron and proton densities ratio.  An example of
this is shown and reviewed.

\subsection{Halo factor}

The simple expression for the antiproton absorption width for the
antiproton which occupies a~state $(s)$ of an atom is
\begin{equation}
  \Gamma_{n(p)}^s \sim \int \,\rho_{n(p)}\,|\Psi^s(r)|^2\,P(r)\, r^2\,dr\,.
  \label{gamma}
\end{equation}
The subscripts $n$ and $p$ are for neutrons and protons respectively.
The wave function $\Psi^s(r)$ of antiproton is determined in the
Schr\"odinger model of hydrogen--like antiprotonic atom within a~point
nucleus. The approximation of the point like nucleus is motivated by
the fact of a~large mean distance $B$ of the antiproton from the
nucleus.  Taking the mass ratio of the antiproton to the electron
$m_{\bar p}/m_e\sim 2000$, the main (supposed) quantum number $n$ of
antiproton in the atom {\it eg.}, $n=8$, and the charge of the nucleus
in question, is $B \approx n^2\,B_0/2000Z$, where $B_0$, is the Bohr
radius of the electron in the hydrogen atom.  For the cases which we
are considering here this gives an average antiproton-nucleus distance
of the value $B \approx 30-40$fm.  As it is seen, this distance exceeds
a few times past the nuclear radius $R\approx A^{1/3}$fm. This
motivates our point nucleus approximation.  It is possible to solve the
Dirac or the Schr\"odinger equation for the more complicated case of
finite or even deformed nucleus. However, this will not change
significantly the relative measures which we shall introduce and apply
in this present paper.

The factor $P(r)$, in the definition of $f$ describes pion escape
probability plus other effects and it will be discussed later.  Here, we
may say only, that in the present model it does not influence the
calculation and with very high accuracy it may be assumed as $P(r)
\approx 1$.

We define the halo factor as~\cite{BP95}
\begin{equation}
  f \sim \frac{Z}{N} \frac {\sum_s\Gamma_n^s} {\sum_s\Gamma_p^s} \,.
  \label{halof}
\end{equation}
A similar definition was used in~\cite{jas+93,lub+94}.  The summation
run through all the antiprotonic atomic states for which one expects
the large annihilation widths. Usually it is assumed that only one
state with $(n,l)$ ranging from $(6,5)$ for Ni nucleus to $(9,8)$ to U
nucleus contributes to these widths~\cite{wycech95}.  In the present
calculation we sum a~few (three to five) states only. The summation
stops if the addition of the successive $\Gamma^s$ term does not change
the value of the sum.

It is worthwhile to mention the weak dependence of our $f$ definition
on the structure of the antiproton-nucleon ($\bar pN$) interaction. It
is assumed that both proton and neutron $\Gamma^s$ widths depend on the
imaginary part of the complex optical potential W(r) which is
responsible for the absorption of the antiproton.  We use the
approximation that the optical potential is proportional to the density
of nucleons: $W(r)=a\,\rho(r)$. Here the parameter $a$ is the $\bar pN$
interaction length and is taken as $a=a_r+ia_i$, where both real $a_r$
and imaginary $a_i$ parts of $a$ are constants~\cite{batty}. In this
way we obtain
\begin{equation}
   \Gamma_n(p) \sim \Im m\,a_{n(p)}\,\rho(r)\,.
\end{equation}
Because both $\sum\Gamma_n^s$ and $\sum\Gamma_p^s$ in the definition of
halo factor $f$ (Eq.~\ref{halof}) appear in a~ratio and thus depend
weakly on the specific antiprotonic atomic state one concludes that the
optical potential dependence of $f$ is washed out.  In this sense $f$
is the $\bar pN$ interaction independent measure of the annihilation
ratio and describes rather the geometrical properties of the nucleon
density distributions.

In fact, the definition of $f$ consists of $\Im m \, a_n/\Im
m\, a_p$ which measures the cross sections annihilation
ratio $\sigma(\bar pn)/\sigma(\bar pp)$ for the antiproton annihilation
on nucleons in a~nuclear medium. The value of this ratio for a~given
nucleus is very hard to estimate in an experiment or on theoretical
grounds.  In an early paper on nucleon--antinucleon optical potential
by Bryan and Philips~\cite{BP68}, the relative antiproton $\bar
p$-capture rates on neutrons and protons in singlet and triplet states
was estimated from measurements of $\bar p$ annihilation at rest in
hydrogen and deuterium. From these one can calculate the ratio of
capture on neutrons to the capture on protons. In singlet states it is
close to 0.6 whereas in triplet states it exceeds 1.  The measurements
by Bugg et al.~\cite{bug+73} show that this ratio is close to 0.65 in
case of the nucleus $^{12}$C.  Other authors \cite{biz+74} give us an
imaginary potential ratio the value $\sim80$\%.  In the paper by
Kalogeropoulos and Tzanakos~\cite{KT76}, one can find a~similar value
$\sigma(\bar pn)/\sigma(\bar pp)=3/4$ as measured in deuterium.  These
different numbers suggest that the annihilation ratio may change from
one nucleus to the other and in our opinion, it is not possible to
scale the results using these numbers -- the ratio may show
a~dependence on proton and neutron numbers $Z$ and $N$ as well as the
state in which the annihilation takes place.

Therefore, in the calculations which follow we do not scale our results.
As we shall see the results show some similarity to the experimental
data.  This supports our choice of the RMF model calculation and shows
that the ratio $\Im m\, a_n/\Im m\, a_p$ is close to unity for our
model.

To select the peripheral annihilation acts (which one observes in $\bar
pX$ experiment) one should correct the definition of the halo factor on
the pion escaping probability $P_{\pi, esc}(r)$ and the deep hole
creation probability $P_{dh}(r)$. One has to take into account only
this part of the annihilation width $\Gamma$ which is responsible for the
cold annihilation --- the annihilation in which pions escape to infinity
without exciting the rest of the system.  On pure geometrical grounds
this is shown to be a~power function of $r$ which agrees roughly with
very accurate calculations presented in Ref.~\cite{wycech95}.  Assuming
the dependence $P_{\pi,esc}\sim r^k$, where $k>0$, one can see from the
Eq.~\ref{halof} that the summation of widths $\Gamma$ will run over
nearly the same spectrum of powers of $r$ as without this factor. It is
therefore allowed to stay with the old formulas in which $P(r)$ is
constant.

Another factor which probably enters the expression for $\Gamma$ is a
deep hole creation factor $P_{dh}(r)$~\cite{wycech95}. Its role is
again the preservation of cold acts of annihilation. The physics behind
this is the following. A~number of annihilation acts may occur on a
deeply situated nucleon levels. This leads to a~rearrangment of the
nucleonic orbits and it may happen that the final system will show very
small or positive(!) binding energies of the last nucleons.  After its
emission, the nucleus in question goes out of the observation range and
can not be counted as the peripheral halo product. The dependence of
the function $P_{dh}$ on $r$ seems to satisfy the power law analogous
to pion escaping probability $P_{\pi,esc}\sim r^k$, and as it was said
before, we do not introduce this into our calculation.  These are all the
approximations which we have assumed in our paper while evaluating the
halo factor.

\subsection{Average densities}

Consider an average density $\bar\rho$ defined by the integral
\begin{equation}
  \bar\rho=\frac{1}{N} \, \int\,d\tau\,\rho^2(r)\,,
  \label{rhoav}
\end{equation}
where $\rho(r)$ is the sphericaly symmetric nucleon density
distribution and $d\tau$ in the integration volume element.  N is the
number of particles in the system and is given by
\begin{equation}
  N=\int\, d\tau\, \rho(r)\,,
\end{equation}
In the case of
pure Fermi distribution
\begin{equation}
  \rho(r)=\rho_0\,\left(1+\exp\frac{r-R}{a}\right)^{-1}\,,
  \label{fermi}
\end{equation}
the quantity $\bar\rho$ can be evaluated approximately with assumed
accuracy.  In the above formula, $a$ is the width of the nuclear
surface, $R$ is the radius and $\rho_0$ is the central density of the
nucleus.  The expression valid to second order in expansion parameter
$a/R$ can be obtained from the Elton's formula (pages 106-107 of
Ref.~\cite{elton61}). To use this formula one has to modify the
expression for the average density to the following form
\begin{equation}
   \bar\rho \sim F_2(k)-2F_1(k)\,,
\end{equation}
where $k=R/a$ and $F_n(k)$ the Elton's integral is defined by
\begin{equation}
   F_n(k)= \int _0 ^\infty \, \frac{x^n \, dx}{(1+\exp(x-k))} \,.
\end{equation}
From the formula mentioned before, one has to second order in $a/R$
\begin{equation}
  \bar\rho \approx \rho_0 \, (1-3 \frac{a}{R}
    + \frac{\pi^2}{2} ( \frac{a}{R} )^2 )\,.
  \label{dfermi}
\end{equation}
A~relatively large, linear term $-3a/R$, which appears in this
expression makes the $\bar\rho$ a~sensitive indicator of the
peripheral properties of the nuclear densities.

On the basis of both average densities $\bar\rho_n$ and
$\bar\rho_p$ one can define the deviation of their ratio from the
homogeneous densities ratio. This is
\begin{equation}
   h=1-\frac{Z}{N}\,\frac{\bar\rho_n}{\bar\rho_p} \,.
   \label{devh}
\end{equation}
Inserting (\ref{dfermi}) into the last equation shows
\begin{equation}
  h \approx 3 \frac{a_n-a_p}{R}\,,
  \label{approxh}
\end{equation}
where $a_n$ and $a_p$ are the diffusness parameters of the neutron and
proton distributions respectively and we have assumed $R_n \approx R_p
= R$.

Figure \ref{figliq} shows an example of the deviation $h$ as calculated
for all of the viewed nuclei in the case of the liquid droplet model
densities~\cite{MS74}. The deviation $h$ was calculated from the
parameters of this model and, the geometric dependencies of the $a$ and
$R$ parameters entering the Fermi density distribution (see
Eq.~\ref{fermi}) and the corresponding parameters of the liquid drop
distribution~\cite{myers73,MS83,HM88}. Figure \ref{figliq} shows two kinds of
data. The impulses show exact (full lines) and approximate (dashed lines)
results respectively. One can see the good agreement of both exact,
and approximate, calculated from Eq. (\ref{devh})
of the density deviation $h$ values.  Large positive
values of $h$ can be seen in the case of neutron rich nuclei like Te, Yb, Th and U.
The deviation $h$ is small and positive in cases of neutron deficient
nuclei: Ni, Zr, Ru.  Latter we shall see a~similar behaviour of $h$
for more realistic calculations done with the RMF model (see Figure
\ref{figdev}).

\section{Relativistic Mean Field Model}

In this section we describe very shortly the relativistic mean field
(RMF) model which we have used in the calculations of the neutron and
proton densities entering both halo factor $f$ and the average density
deviation.

The RMF theory starts from a~lagrangian consisting of nucleonic and
mesonic degrees of freedom \cite{SW86}.  In some sense it seems to be
more fundamental than other mean field models like Skyrme-Hartree-Fock
and Gogny-Hartree-Fock models.  It gives the relativistic treatment of
nucleonic and mesonic variables and a~proper description of the
spin-orbit interactions.  Nevertheless it is still an effective
phenomenological method based on the local densities and fields.

The RMF theory was successfully used to reproduce parameters of the
nuclear matter and some properties of finite nuclei like binding
energies, mean square charge radii and quadrupole
moments~\cite{GRT90,bar+95}.

This theory is based on the following field lagrangian density
\cite{SW86,GRT90,HS81,HS83,Bou82}
\begin{eqnarray}
{\cal L}&=& \bar\psi_i\{i\,\gamma^\mu\,\partial_\mu - M\}\psi_i
  \nonumber \\
  &&+\frac{1}{2}\partial^\mu\sigma\,\partial_\mu\sigma 
    - \frac{1}{2} m_\sigma\sigma^2
    -g_\sigma\,\bar\psi_i\psi_i\,\sigma \nonumber \\
  \label{LRMF}
  &&-\frac{1}{4}\Omega^{\mu\nu}\Omega_{\mu\nu} 
    +\frac{1}{2}m_\omega^2\,\omega^\mu\omega_\mu 
    -g_\omega\,\bar\psi_i\,\gamma^\mu\,\psi_i\,\omega_\mu \\
  &&-\frac{1}{4}\vec{R}^{\mu\nu}\vec{R}_{\mu\nu} 
    +\frac{1}{2}m_\rho^2\,\vec\rho\,^\mu\vec\rho_\mu 
    -g_\rho\bar\psi_i\,\gamma^\mu\vec\tau\,\psi_i\,\vec\rho_\mu
    \nonumber \\
  &&-\frac{1}{4}F^{\mu\nu}F_{\mu\nu} 
    -e\bar\psi_i\,\gamma^\mu\,\frac{(1-\tau_3)}{2}\,\psi_i\,A_\mu\;.
  \nonumber
\end{eqnarray}
The fields belong to nucleons (Dirac spinor field $\psi$), the low mass
isovector-vector meson $(\vec\rho_\mu; \vec{R}_{\mu\nu})$,
isoscalar-vector $(\omega_\mu; \Omega_{\mu\nu})$, scalar $\sigma$ and
to the massless photon vector field $(A_\mu; F_{\mu\nu})$.  The field
tensors are given by
\begin{equation}
\Omega^{\mu\nu} = \partial^\mu\omega^\nu - \partial^\nu\omega^\mu \,\, ,
\end{equation}
\begin{equation}
\vec{R}^{\mu\nu} = \partial^\mu\vec\rho\,^\nu - \partial^\nu\vec\rho\,^\mu
                -g_\rho (\vec\rho\,^\mu \times \vec\rho\,^\nu)\,\, ,
\end{equation}
\begin{equation}
  F^{\mu\nu} = \partial^\mu A^\nu - \partial^\nu A^\mu \,\, .
\end{equation}

The Dirac spinors $\psi_i$ of the nucleon and the fields of
$\sigma,\,\rho$ and $\omega$ mesons are solutions of the coupled Dirac
and Klein--Gordon equations which are obtained from Eq.~\ref{LRMF} by
means of the classical variational principle and they are then solved
by iteration for the case of the spherically--symmetric systems of
nucleons.  This iteration goes through the following steps: we start
from an estimate of meson and electromagnetic fields and we solve the
Dirac equation getting the spinors $\psi_i$. The spinors are then used
to obtain the densities. The latter serve as source to solve the Klein
Gordon equations and achieve the new estimations of the meson and
electromagnetic fields.  The parameters used in our calculation are the
same as in Refs.~\cite{HS81,HS83,HMS91}.

\section{Results}

On the basis of the relativistic mean field model which was presented
in the previous section and is described in detail in
\cite{SW86,HS81,HS83,HMS91,Bou82,GRT90} we have calculated the nucleon
densities entering the Eq.~\ref{gamma} for the absorption width and the
halo factor Eq.~\ref{halof}.  We show that some of the nuclei are good
candidates for the neutron halo systems and have promising large
logarithm of neutron to proton density ratio $l(r)$.  This ratio which
we define as
\begin{equation}
   l(r)=\log(\rho_n(r)/\rho_p(r))\,. \label{logf}
\end{equation}
is a~sensitive quantity and it suits to relate densities especially in
a peripheral nuclear region where both neutron and proton densities are
very small.  Figure \ref{figlog} shows this ratio for all nuclei under
consideration.

In Figure \ref{figlog} one can see nuclei for which the logarithm
$l(r)$, for $r>2R$, has the value of few orders.  This means that the
neutron density is there still greater then the proton density.  This
indicates a~possible large halo factor $f$.

%%% KP

One has to remember that the microscopic density $\rho$ is the sum of
the single particles densities $\rho_\nu$ over all occupied states
\begin{equation}
    \rho(r)=\sum_{\nu, {\rm occ}} \rho_\nu(r)\,.
    \label{rho}
\end{equation}
It is interesting to study the dependence on $r$ of the contribution of
the single particle orbitals $\nu=(nlj)$ to the total density?  Or in
another words what is the collectivity of the nuclear density tail?
The results of the RMF calculation for some nuclei in which the halo
factor is discussed are presented in Figure~\ref{figtails}.  It is seen
that for large distances of 10--14fm only a~few valence orbitals
contribute to the density tail.  Even more, for some nuclei in which
$f$ is large, one neutron orbital exhausts 90\%\ of the nuclear
density. Only in the case of $^{144}$Sm in addition to the neutron
orbital $1h_{9/2}$ there appears $2d_{5/2}$ proton orbital with the
comparable amplitude. In this nucleus one really observes very small
neutron halo factor $f$ (proton halo).

We have to add from our numerical experience that the asymptotic
behaviour of the single particle density depends dramatically on the
single particle energy. For the orbitals less bounded the large $r$
tail of density is longer. So the halo effect will not be the only
probe of the surface width of the neutron and proton distributions but
also a~crucial test of the single particle structure foreseen by
theoretical models. This also means the inclusion of a~nuclear shape
deformation into our model could be important.

%% KP end

Figure \ref{figlogro} shows the dependence of the logarithm of the
total density of baryons against the distance $r$ measured from the
center of the nucleus.  The logarithm of the density in the peripheral
area of the nucleus ($r>8$) takes very small values and is nearly
linear function of $r$ with a~steep negative slope.  All the slopes for
different nuclei are similar.

We now consider another two quantities which characterize the real
density distribution. The first is the deviation of density ratio $h$
from the homogeneous density ratio (\ref{devh}).  In contrary to the
case shown already in Figure \ref{figliq}, this is calculated for real
density distributions determined from the RMF model.  In analogy to
Eq.~\ref{devh}, it is defined through the average densities calculated
in RMF model. The second quantity we considered is the celebrated halo
factor $f$ (Eq. \ref{halof}).  Both quantities give the measure of the
density distribution in the nucleus.  These are shown in
Figure~\ref{figdev} and Figure~\ref{figfac} respectively.

One can easily identify a~halo nuclei. In Figure \ref{figdev} you can
see an interesting case of Ru nucleus for which the relative density
deviation $h$ is negative. This suggests the lack of the neutron halo
for this nucleus.  At the same time the halo factor $f \approx 2$. We
concluded that the deviation $h$ is rather a~rough indication of the
neutron halo in nuclei.

The calculated halo factors $f$ are shown in Figure~\ref{figfac} and in
Table~1 where in addition we have also displayed the experimental
values $f_{\rm exp}$. These values were extracted from
Ref.~\cite{lub+94}.  Except for extreme cases of $^{96}$Ru, $^{144}$Sm
and $^{176}$Yb one can observe a~good agreement between calculated
factor $f_{\rm RMF}$ and the experimental data.  The case of $^{176}$Yb
it is a~special one. It shows a~too low theoretical halo factor $3.1$
and a~very high measured value of $8.1$. This is the case discussed
also in \cite{wycech95}. The case of samarium nucleus, $^{144}$Sm,
shows $f$ which probably indicates the {\bf proton halo} instead of
neutron one. In our calculations we obtained in this case $f=1.5$.  The
similar situation of large theoretical $f$ is observed in the nucleus
$^{96}$Ru.

%%%%%%%%%%%%%%%%%%%

\section{Summary}

In the relativistic mean field model we have calculated the halo factor
$f$ and the factor $h$ -- the average density ratio deviation from the
ratio of homogeneous density distributions, for nuclei in which we
observed the neutron halo.

Both quantities $f$ and $h$ indicate the possible halo effects in most
of the studied nuclei but in some cases (see eg., Yb nucleus) they
differ from experimental data significantly. This fact shows that one
has to extend the RMF nuclear density model. The extension may include
{\it eg.}, the deformation of the nuclear systems or/and the pairing
interaction. The finite deformation in all considered cases may change
the picture a~little bit of the neutron halo.

The new parameter characterizing the peripheral properties of the
nuclear density distributions, which we have called the average density
ratio deviation $h$, is proportional to the difference of the proton
and the neutron surface diffuseness parameters.  Like $f$, it points
out these nuclei which show the neutron halo properties.

We have shown that the nuclear density tail manifests a~single particle
nature. Therefore, the neutron halo is the single particle effect and
not a~collective one.

Some of our predictions, as compared to the experimental data, fail but
most of them show correlations which are very promising ones.  It is
hard to explain the source of existing divergences. A~possible
explanation may be the lack of deformation in our model calculations.
It is also possible that the inclusion of pairing interaction to this
theory may improve the results. Such calculations are in progress.

%%%%%%%%%%%%
\acknowledgements
We cordially acknowledge the discussions with Professor
J.~Jastrz\c{e}bski from the Warsaw University and Professor S. Wycech
from IPJ, Warsaw.

%%%%%%%%%%%%%%%%%%%
\clearpage
\bigskip\bigskip

\begin{center}
\begin{table}
\caption{\bf Experimental and calculated halo factors.}
\begin{tabular}{c|c|c}
Nucleus         & $f_{\rm exp}$ & $f_{\rm RMF}$ \\
\hline 
$^{58}$Ni       &       $0.9$   &       $1.2$   \\
$^{96}$Zr       &       $2.6$   &       $2.3$   \\
$^{96}$Ru       &       $0.8$   &       $2.3$   \\
$^{130}$Te      &       $4.1$   &       $3.5$   \\
$^{144}$Sm      &       \underbar{$<0.4$}       & $1.5$ \\
$^{154}$Sm      &       $2.0$   &       $3.0$   \\
$^{176}$Yb      &       \underbar{$8.1$}        &  $3.6$        \\
$^{232}$Th      &       $5.1$   &       $5.5$   \\
$^{238}$U       &       $6.0$   &       $5.0$   \\
\end{tabular}
\end{table}
\end{center}

%%%%%% FIGURES %%%%%%%%%%% FIGURES    %%%%%%%%%   FIGURES

\clearpage

\begin{figure}\caption{\label{figliq}
The deviation $h$ of the neutron to proton density ratio for Fermi
density distributions in the case of liquid droplet model. The impulses
show exact (full line) and approximate (dashed line) results.
The data corresponds to the following nuclei: $^{58}$Ni, $^{96}$Zr,
$^{96}$Ru, $^{130}$Te, $^{144}$Sm, $^{154}$Sm, $^{176}$Yb, $^{232}$Th,
$^{238}$U.}
\end{figure}

\begin{figure}\caption{\label{figlog}
The logarithm of the neutron to proton density ratio calculated on the
basis of relativistic mean field model.  The $l(r)$ curves are shifted
on a value $\delta$ given on the right hand side of each curve.}
\end{figure}

\begin{figure}\caption{\label{figtails}
The partial density ratio for single particle neutron (n; full line)
and proton (p; dashed line) orbitals. Arrows show the locis of
the nuclear root mean square radii of the charge distributions.}
\end{figure}

\begin{figure}\caption{\label{figlogro}
The logarithm of the baryon density against the distance $r$ (in fm)
measured from the center of the nucleus.}
\end{figure}

\begin{figure}\caption{\label{figdev}
The deviation $h$ of the neutron to proton density ratio.  The nuclei
are the same as in Figure 2.}
\end{figure}

\begin{figure}\caption{\label{figfac}
Halo factor $f$. The nuclei are the same as in Figure 2. The experimental 
data (squares) were taken from \protect\onlinecite{lub+94}.}
\end{figure}

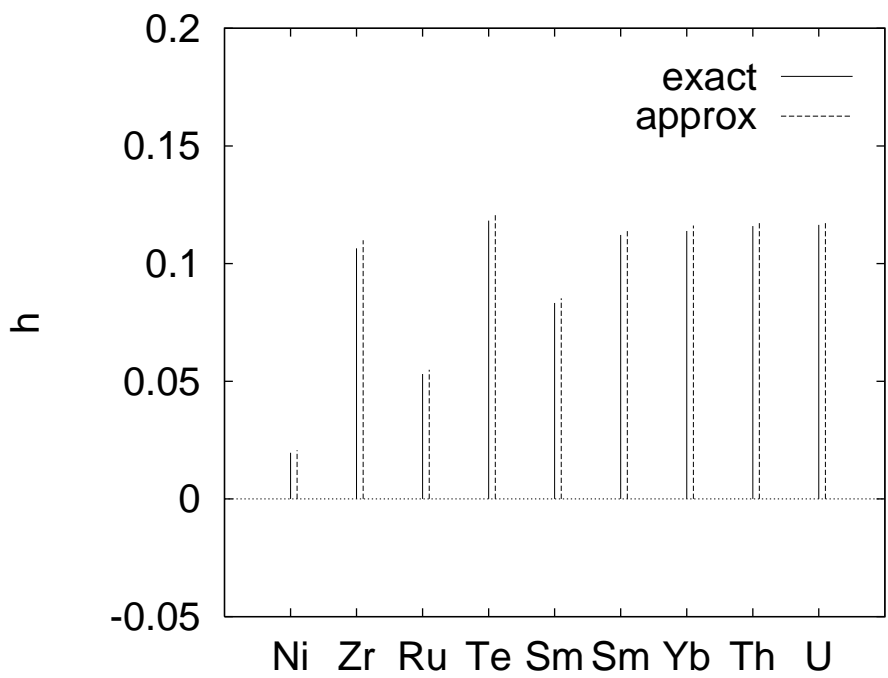
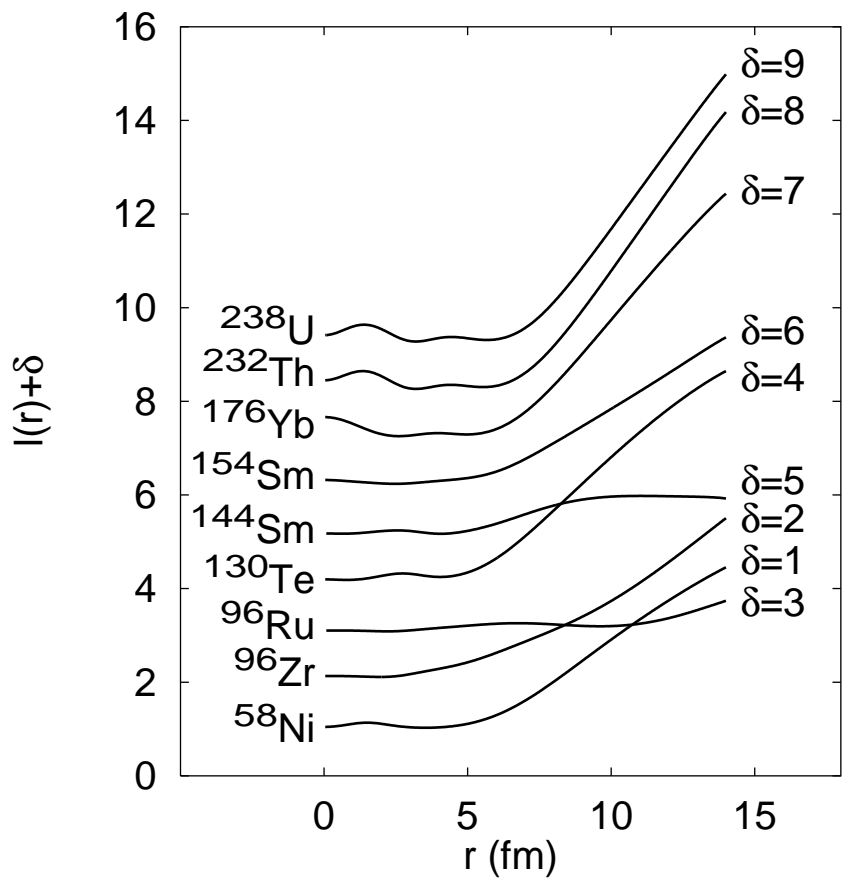
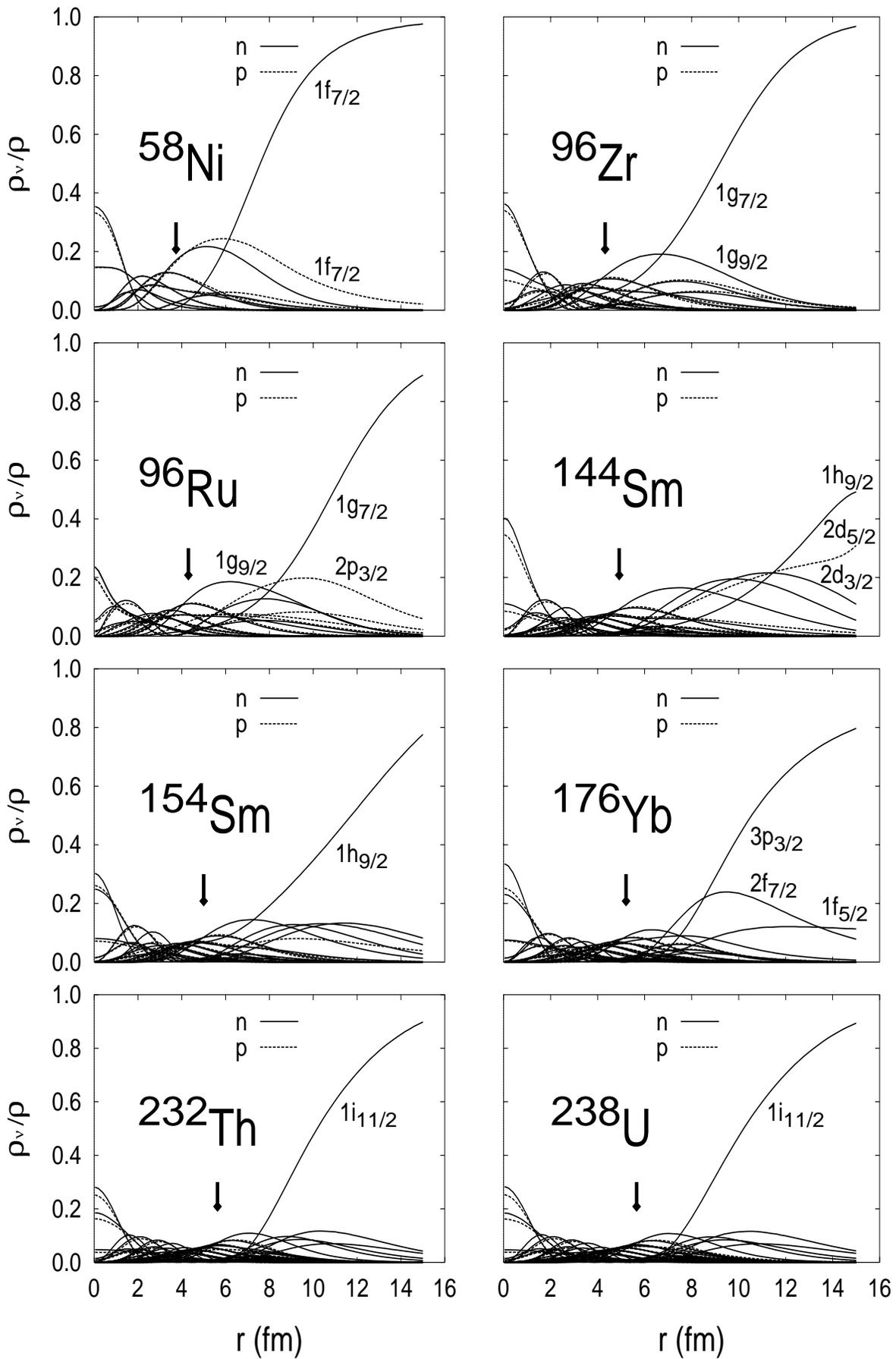
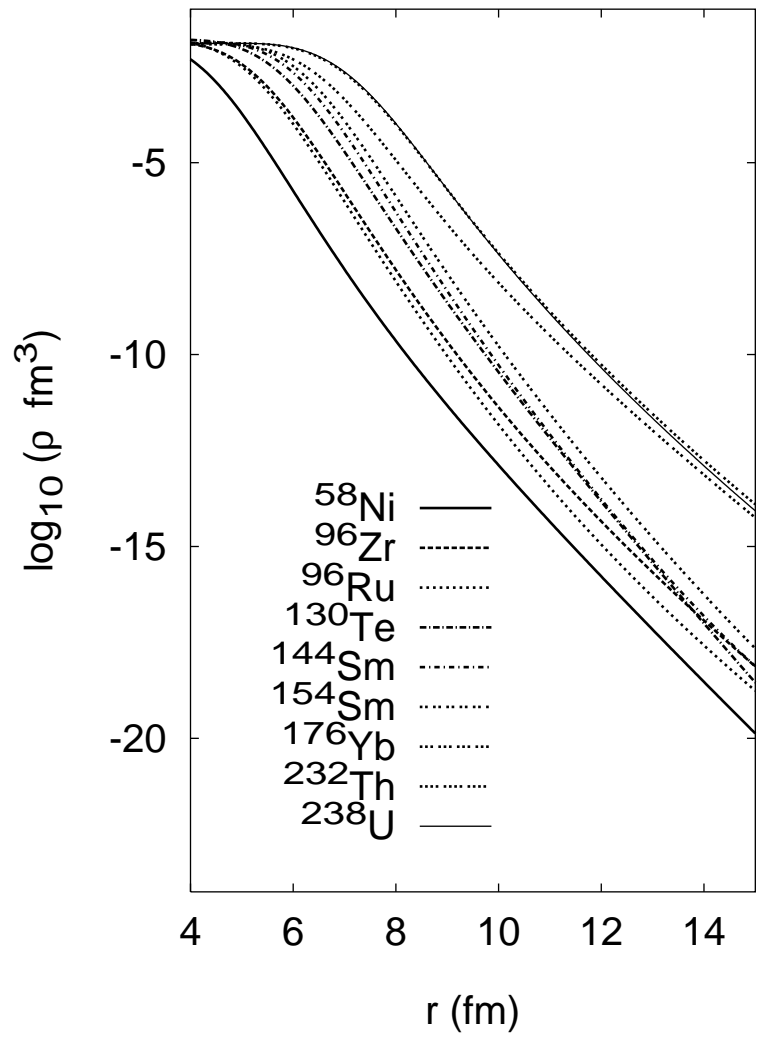
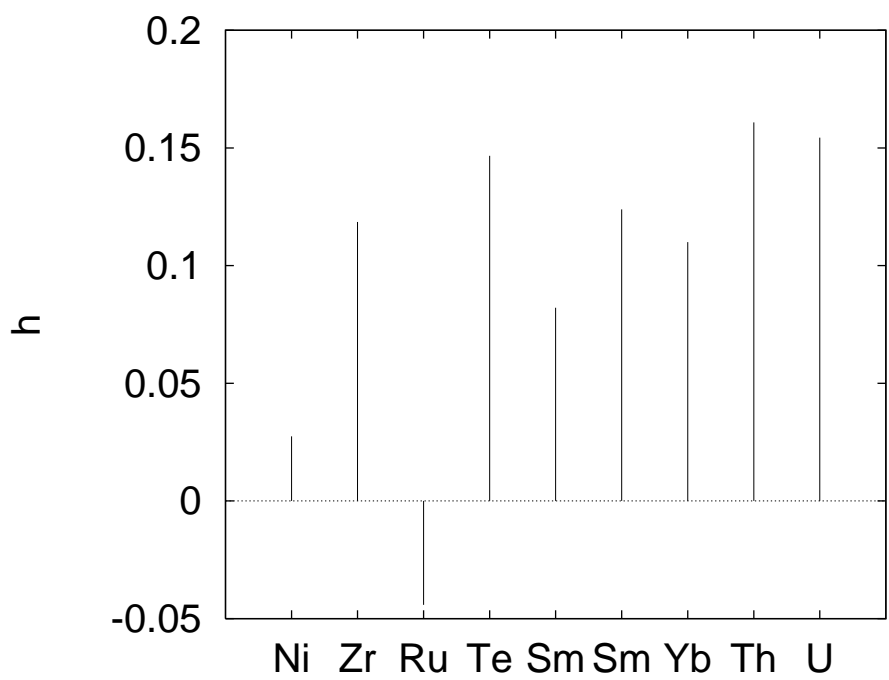
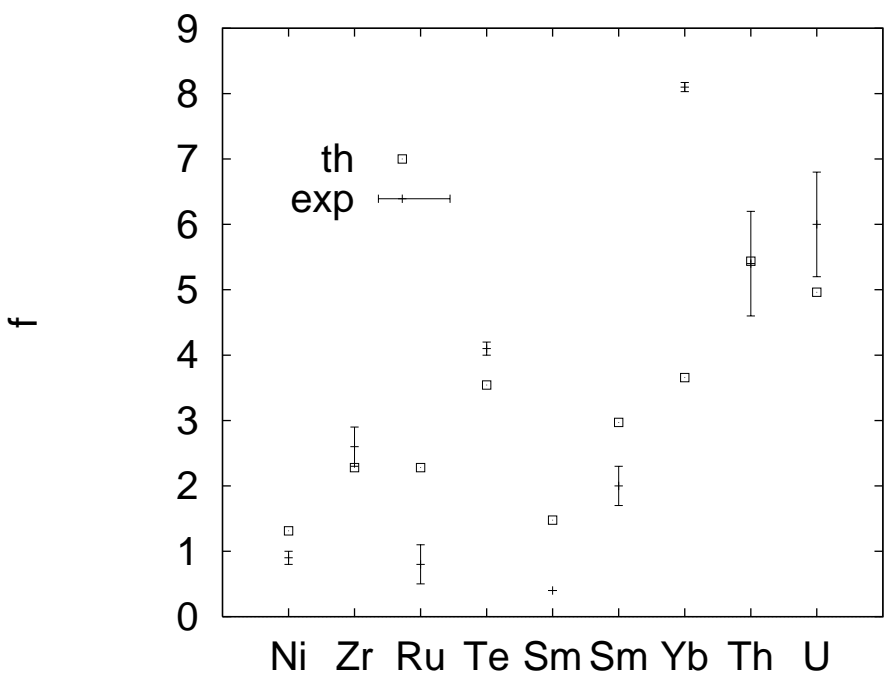
\end{document}